\documentclass[12pt]{article}
\usepackage{graphicx}
\def\mydate{28 April 2005}

\def\ignore#1{{}}

\let\oldtheequation=\theequation
\def\doteqs#1{\setcounter{equation}{0}            
\def\theequation{{#1}.\oldtheequation}}
\newcounter{sxn}
\def\sx#1{\addtocounter{sxn}{1} \vskip 0.8cm  \goodbreak
\noindent{\large\bf\leftline{\thesxn.~~#1}} \nobreak \vskip -.5cm}
\def\sxn#1{\sx{#1} \doteqs{\thesxn}}

\newcounter{axn}

\date{}

\newdimen\mybaselineskip
\newcommand{\beeq}{\begin{equation}}
\newcommand{\eneq}{\end{equation}}
\newcommand{\beqn}{\begin{eqnarray}}
\newcommand{\eeqn}{\end{eqnarray}}

\def\mybig{\displaystyle \strut }

\def\dd{\partial}
\def\la{\raise.16ex\hbox{$\langle$}\lower.16ex\hbox{}  }
\def\ra{\, \raise.16ex\hbox{$\rangle$}\lower.16ex\hbox{} }

\def\onehalf{ \hbox{${1\over 2}$} }

\def\eff{{\rm eff}}
\def\min{{\rm min}}
\def\sym{{\rm sym}}

\def\tr{\,{\rm tr} \, }

\def\psibar{ \psi \kern-.65em\raise.6em\hbox{$-$} \lower.6em\hbox{} }
\def\psibarb{ \psi \kern-.65em\raise.6em\hbox{$-$}  }

\def\myfrac#1#2{{\mybig #1\over \mybig #2}}

\def\boxit#1{$\vcenter{\hrule\hbox{\vrule\kern3pt
     \vbox{\kern3pt\hbox{#1}\kern3pt}\kern3pt\vrule}\hrule}$}
\def\bigbox#1{$\vcenter{\hrule\hbox{\vrule\kern5pt
     \vbox{\kern5pt\hbox{#1}\kern5pt}\kern5pt\vrule}\hrule}$}
\def\Bigbox#1{$\vcenter{\hrule\hbox{\vrule\kern7pt
     \vbox{\kern7pt\hbox{#1}\kern7pt}\kern7pt\vrule}\hrule}$}

\begin{document}

\thispagestyle{empty}

\baselineskip=25pt

{\small \noindent \mydate \hfill OU-HET 520/2005}
\vspace*{1.cm}

\begin{center}  
{\Large \bf Dynamical Gauge Symmetry Breaking} \\
{\Large \bf by Wilson Lines in the Electroweak Theory\footnote{To appear 
in the proceedings of 
{\it ``International Workshop on Dynamical Symmetry Breaking''} (DSB 2004), 
Nagoya University, Nagoya, Japan, December 21-22, 2004.}} \\
\end{center}

\baselineskip=14pt

\vspace{.5cm}
\centerline{\bf  Yutaka Hosotani}

\centerline{\small \it Department of Physics, Osaka University, Toyonaka, 
Osaka 560-0043, Japan}

\vspace{.5cm}
\begin{abstract}
In higher dimensional gauge theory, dynamics of non-Abelian Aharonov-Bohm 
phases induces gauge symmetry breaking through the Hosotani mechanism.
Higgs fields in the four-dimensional spacetime are identified with 
the extra-dimensional components of the gauge fields.   Basics of the 
Hosotani mechanism are reviewed, and applied to the electroweak theory.
The Higgs boson mass and the Kaluza-Klein excitation scale are related
to the weak $W$-boson mass.
\end{abstract}

\baselineskip=17pt plus 1pt minus 1pt

\vskip .2cm 

\sxn{Introduction }
Gauge symmetry breaking is at the core of the current understanding of
the particle interactions. Yet the Higgs particle remains as an enigma 
in the unified electroweak theory.  Does it really exist?  How heavy is 
it if it exists?
How does it interact with quarks and leptons?  These are the issues 
to be settled in the forthcoming experiments at LHC.  In the standard model 
of electroweak interactions, the mass of Higgs bosons is in large part 
unconstrained.  In the minimal supersymmetric standard model (MSSM) 
the mass of the lightest Higgs boson is predicted in the range 
100$\,$GeV to 130$\,$GeV.  The experimentally preferred value is
$126^{+73}_{-48} \,$GeV.  In this lecture we explore an alternative
scenario, the dynamical gauge-Higgs unification, to try to pin down 
the nature of Higgs bosons.

In the dynamical gauge-Higgs unification formulated in a gauge theory
in higher dimensions,  extra-dimensional components of
gauge fields play the role of Higgs fields in the four-dimensional 
spacetime.\cite{Witten1,Fairlie1,Manton1}
When the extra-dimensional space is not simply connected, there appear
non-Abelian Aharonov-Bohm phases, or Wilson line phases, 
whose fluctuation modes in the four dimensions serve as Higgs scalar
fields.  They are massless at the tree level. Its effective 
potential is completely flat at the classical level in the directions
of Aharonov-Bohm phases, but becomes non-trivial at the quantum level.
They may develop non-vanishing expectation values, thus inducing
dynamical gauge symmetry breaking.  This is called the Hosotani 
mechanism.\cite{YH1,YH2}

We first review the Hosotani mechanism to see how dynamics of Wilson line
phases induce gauge symmetry breaking.  Examples are given in $SU(N)$ gauge
theory on $M^n \times S^1$.  Then we explain how the scenario can be implemented
in electroweak interactions by considering gauge theory on orbifolds.
Detailed analysis is given in the $U(3) \times U(3)$ model on 
$M^4 \times (T^2/Z_2)$.  The mass of the Higgs boson and the Kaluza-Klein 
mass scale is determined.  

\vskip 1.5cm

\leftline{\bf \large Part I.  Dynamical gauge symmetry breaking by Wilson lines}

\sxn{$SU(N)$ gauge theory on $M^4 \times S^1$}

If the space is not simply connected, Wilson line phases become
physical degrees of freedom.  Although constant Wilson line phases
yield vanishing field strengths, they are dynamical and affect physics.
At the classical level Wilson line phases label degenerate vacua.
The degeneracy is lifted by quantum effects. The effective potential of 
Wilson line phases become non-trivial.   If the effective potential is 
minimized at nontrivial values of Wilson line phases, then the 
rearrangement of gauge symmetry takes place. Spontaneous
gauge symmetry breaking or enhancement is achieved 
dynamically.

Let us take $SU(N)$ gauge theory on $M^4 \times S^1$ as an example.
Let $x^\mu$ and $y$ be coordinates of $M^4$ and $S^1$, respectively.
Points $y$ and $y + 2\pi R$ are identified on $S^1$.  Gauge theory is
defined first on the covering space of $M^4 \times S^1$, namely 
on $M^4 \times R^1$, on which all fields are smooth. On $S^1$,
physics must be the same at $y$ and $y + 2\pi R$.  However, it does
not necessarily means that fields themselves are the same.  Upon
a loop translation along $S^1$, each field needs to come back to
the original value up to a (global) gauge transformation.  
\beqn
&&\hskip -1cm 
A_M(x,  y +2 \pi R) =
U A_M(x,  y ) \, U^\dagger  ~~~, \cr
\noalign{\kern 5pt}
&&\hskip -1cm 
\psi(x, y + 2 \pi R) 
  =  e^{i\beta} \, T[U] \psi(x,  y) ~~~~.
\label{BC1}
\eeqn
$U$ is an element of $SU(N)$.   $T[U] \psi= U \psi$ or   
$U \psi U^\dagger$ for $\psi$ in the fundamental or adjoint representation,
respectively. The boundary condition (\ref{BC1}) guarantees that the physics is 
the same at $(x,  y)$ and $(x,  y + 2 \pi R)$.
The theory is defined with a set of boundary conditions $\{ U, \beta \}$. 

One might ask the following questions. Does $U \not\propto I$ imply 
symmetry breaking?  What is the 
symmetry of the theory for a general $U$?  Answers to these questions are
quite nontrivial.  Under a gauge transformation
\beeq
A_M'
=\Omega \bigg( A_M -{i\over g}\partial_M\bigg) \Omega^{\dagger} ~~, 
\label{gaugeT1}
\end{equation}
$A_M'$ obeys a new set of boundary conditions 
\beqn
&&\hskip -1cm 
A_M'(x,  y +2 \pi R) =
U' A_M'(x,  y ) \, U'^\dagger  ~~~, \cr
\noalign{\kern 5pt}
&&\hskip -1cm
U' = \Omega(x,  y + 2 \pi R)\, U \, \Omega(x,  y)^\dagger ~, 
\label{BC2}
\eeqn 
provided  $\dd_M U' = 0$.
The set $\{ U' , \beta \}$ can be different from the set $\{ U, \beta \}$.  
When the relation $\dd_M U' = 0$ is satisfied, we write
\beeq
\{ U' , \beta \} ~\sim~ \{ U, \beta \} ~~.
\label{relation1}
\eneq
The relation is transitive, and therefore is an equivalence
relation.  Sets of boundary conditions form   equivalence classes of boundary 
conditions with respect to the equivalence 
relation (\ref{relation1}).\cite{YH2}  As an example we note
\beqn
U = \pmatrix{ e^{i\alpha_1} \cr
& \cdots \cr && e^{i\alpha_N} }
&,&
\Omega  = \pmatrix{ e^{i\gamma_1 y/2\pi R} \cr
& \cdots \cr && e^{i\gamma_N y/2\pi R} }  \cr
\noalign{\kern 10pt}
&\Rightarrow&
U' = \pmatrix{ e^{i(\alpha_1 + \gamma_1)} \cr
& \cdots \cr && e^{i(\alpha_N + \gamma_N)} } ~~.
\label{BC3}
\eeqn
Although the theories defined with 
$\{ U, \beta \}$ and $\{ U', \beta \}$ seem different, they should be 
equivalent and should have the same physics as they are related to each other by
a ``large gauge transformation''.  

The equivalence of physics is guaranteed by the dynamics of Wilson line phases.
Take a theory with the boundary conditions (\ref{BC1}). Given $U$, there 
are zero modes ($(x^\mu, y)$-independent modes) of $A_y$  satisfying
$[ A_y , U ] = 0$.  Although they give vanishing field strengths, 
they cannot be gauged away in general.  Indeed, eigenvalues
$\{ e^{i\theta_1}, \cdots, e^{i\theta_N} \}$ of 
$P \exp \Big( ig \int_0^{2\pi R} dy \, A_y \Big) \cdot U$ are
invariant under all gauge transformations preserving the boundary
conditions (\ref{BC1}). 
$\{ \theta_1, \cdots, \theta_N \}$ are the Wilson line phases. 
They are non-Abelian Aharonov-Bohm phases.

The effective potential for $\{ \theta_1, \cdots, \theta_N \}$ becomes 
nontrivial at the quantum level.  At the one loop level
\beeq
V_\eff [\theta_W]^{d=5} = \sum (\mp) \myfrac{i}{2} \tr \ln D_M D^M 
\label{effV1}
\eneq
where $D_M D^M = \dd_\mu \dd^\mu - D_y^2$.
$D_y$ stands for a covariant derivative with a constant $A_y$ yielding
$\theta_W$'s.  The sign is $-$ ($+$) for a boson (fermion).  Given the 
boundary conditions and background $A_y$, the spectrum of each field is
determined.  On $S^1$ the spectrum in the $y$-direction  takes 
the form $[n+\gamma(\theta_W)]^2/R^2$ where $n$ runs over integers.
Here $\gamma(\theta_W)$ depends on the boundary conditions and couplings
of the fields.  It satisfies that 
$\gamma(\theta_W + 2 \pi) = \gamma(\theta_W) + \ell$ where $\ell$ is an
integer.  Hence, after making a Wick rotation, the four-dimensional 
$V_\eff$ becomes
\beqn
&&\hskip -1cm
V_\eff [\theta_W ] = (2\pi R) \cdot   V_\eff [\theta_W ]^{d=5} \cr
\noalign{\kern 10pt}
&&\hskip .5cm
= \sum (\pm) \myfrac{1}{2} 
\int \myfrac{d^4 p_E}{(2\pi )^4} 
\sum_{n=-\infty}^\infty  \ln \bigg\{ p_E^2 + 
    \myfrac{[n+\gamma(\theta_W)]^2}{R^2} \bigg\} ~~. 
\label{effV2}
\eeqn
As will be proven in the next section, the $\theta_W$-dependent part of 
$V_\eff [\theta_W ]$ is finite. It is given by
\beqn
&&\hskip -1cm
V_\eff [ \theta_W ] = \sum (\mp) \myfrac{3}{64 \pi^6 R^4} ~
 h_5 [\gamma(\theta_W)] + {\rm constant} ~, \cr
\noalign{\kern 10pt}
&&\hskip -.6cm
h_d(x) = \sum_{n=1}^\infty \myfrac{\cos 2\pi n x}{n^d} ~~.
\label{effV3}
\eeqn

The effective potential has a global minimum at $\theta_W = \theta_W^\min$, 
depending on the content of matter fields.   When $\theta_W^\min \not= 0$,
the physical symmetry of the theory differs from the symmetry determined
by the boundary condition matrix $U$ in (\ref{BC1}). To find the physical 
symmetry it is most convenient to make a general gauge transformation
which alters boundary conditions.  Let $A_y^\min$ be the constant gauge 
potential  corresponding to $\theta_W^\min$.  It follows from (\ref{BC1}) that 
$[A_y^\min , U] = 0$.  We perform a gauge transformation (\ref{gaugeT1}) with
$\Omega(y) = \exp \big\{ ig y A_y^\min \big\}$.  In the new gauge
 the boundary condition matrix changes to
$U' = U \exp \big\{2 \pi i g R A_y^\min \big\}  \equiv U^\sym$
 as specified in (\ref{BC2}).
Since the effective potential is minimized at $A_y' = 0$, the physical
symmetry is given by the symmetry specified with  $U^\sym$.

\sxn{Finiteness of $V_\eff [\theta_W ]$}

Although gauge theory in higher dimensions is not renormalizable, 
the $\theta_W$-dependent part of the effective potential can be 
evaluated unambiguously.  It turns out finite at the one loop level, 
being free from divergences which may sensitively depend on physics 
at much higher energy scales.   The $\theta_W$-dependent 
parts of all physical quantities might finite.\cite{Morris}  
In this section 
we show how (\ref{effV3}) is derived,  and  present  theorems.

Consider a quantity
\beeq
f(x) = \myfrac{1}{2}  \int \myfrac{d^d q_E}{(2\pi )^d} 
\sum_{n=-\infty}^\infty  \ln \big\{ q_E^2 + (n+x)^2  \big\} ~~.
\label{formula1}
\eneq
The $x$-dependent part of $f(x)$ is easily found by the zeta
function regularization.  Associated with $f(x)$, $\zeta (s; x)$ is 
defined by
\beeq
\zeta (s; x) = \myfrac{1}{2}  \int \myfrac{d^d q_E}{(2\pi )^d} 
\sum_{n=-\infty}^\infty  \Big\{ q_E^2 + (n+x)^2  \Big\}^{-s} 
\quad {\rm for ~}  \Re  \, s > \myfrac{d+1}{2}  ~.
\label{formula2}
\eneq 
For $\Re \, s \le \onehalf (d+1)$, $\zeta (s; x)$ is defined by analytic
continuation.  $f(x)$ is, then, given by
\beeq
f(x) = - \zeta'(0; x) ~~. 
\label{formula3}
\eneq
Making use of
\beqn
&&\hskip -1cm
\myfrac{1}{A^s} = \myfrac{1}{\Gamma(s)}
\int_0^\infty dt \, t^{s-1} \, e^{-At} ~~, \cr
\noalign{\kern 10pt}
&&\hskip -1cm
\sum_{n=-\infty}^\infty  e^{ - t (n+x)^2 }
= \bigg( \myfrac{\pi}{t} \bigg)^{1/2}
\sum_{n=-\infty}^\infty  
e^{ - \pi^2 n^2/t + 2\pi i nx} ~~,
\label{formula4}
\eeqn
$\zeta (s; x)$ is transformed to
\beqn
&&\hskip -1cm
\zeta (s; x) = \myfrac{1}{\Gamma(s)} \int_0^\infty dt \, t^{s-1} \,
\myfrac{1}{2}  \int \myfrac{d^d q_E}{(2\pi )^d} 
\sum_{n=-\infty}^\infty 
e^{ - [\,  q_E^2 + (n+x)^2 \,  ] \, t } \cr
\noalign{\kern 10pt}
&&\hskip 0.2cm
=  \myfrac{1}{\Gamma(s)} \myfrac{1}{2^{d+1} \pi^{(d-1)/2}}
\int_0^\infty dt \, t^{s-1-(d+1)/2} 
\sum_{n=-\infty}^\infty  
e^{ - \pi^2 n^2/t  + 2\pi i nx} ~.
\label{formula5}
\eeqn
The $n=0$ term in the last line is independent of $x$.  Hence
\beqn
&&\hskip -1cm
f(x) = -  \myfrac{1}{2^{d+1} \pi^{(d-1)/2}}
\int_0^\infty dt \, t^{-1-(d+1)/2} 
\sum_{n \not= 0} 
e^{- \pi^2 n^2 / t + 2\pi i nx  } + {\rm constant} \cr
\noalign{\kern 10pt}
&&\hskip 0.2cm
= -  \myfrac{\Gamma \Big( \onehalf(d+1) \Big)}{2^{d} \pi^{(3d+1)/2}} 
~ h_{d+1} (x)   + {\rm constant}  ~~.
\label{formula6}
\eeqn 
The formula (\ref{formula6}) with $d=4$ leads to (\ref{effV3}). 

$h_d(x)$ is periodic; $h_d(x+1) = h_d(x)$.  It  is, in general, 
singular at $x=$an integer. 
For examples,
\beqn
&&\hskip -1cm
h_2(x) = \pi^2 (x - \onehalf )^2 - \myfrac{\pi^2}{12} ~~, \cr 
\noalign{\kern 5pt}
&&\hskip -1cm
h_4(x) = - \myfrac{\pi^4}{3} (x - \onehalf)^4 
 + \myfrac{\pi^4}{6} (x - \onehalf)^2 - \myfrac{7 \pi^4}{720} ~~,
\label{example1}
\eeqn
for $0 \le x \le 1$.  It is easy to see
\beeq
h_2''(x) = 2\pi^2 \Big\{ 1 - \delta_1(x) \Big\}
\label{example2}
\eneq
where $\delta_a(x) = \sum_n \delta (x- an)$.  For even $d \ge 4$, 
$h_d^{(d-2)} (x) = (2\pi)^{d-2} (-1)^{(d-2)/2} h_2(x)$ so that
$h_d(x)$ has a singularity of cusp type at  $x=n$ ($n$: an integer).

The behavior of $h_d(x)$ for odd $d$ is slightly different.
Recall
\beqn
&&\hskip -1cm
h_1(x) = - \ln \Big( 2 \sin \pi x \Big) \qquad {\rm for ~}
0 < x < 1 ~, \cr
\noalign{\kern 10 pt}
&&\hskip -1cm
h_1'(x) = - \myfrac{\pi}{2} \cot \pi x ~~.
\label{example3}
\eeqn 
Hence $h_d(x)$ has singular behavior at $x=0$ as $x^{d-1} \ln |x|$ for
odd $d$. 

The $\theta_W$-dependent part of the effective potential $V_\eff$
turns out finite.    We summarize it in a theorem.

\bigskip

\noindent
{\bf Theorem}
\vskip .2cm 
\centerline{\bigbox{
\vtop{\hsize=14cm
\noindent
The  effective potential for the Wilson line phases, $V_\eff(\theta_W)$, 
is finite at the one loop level apart from  a $\theta_W$-independent 
constant term.
}}}

\vskip .4cm 
\noindent
{\bf (Proof)} \quad 
In general there are several Wilson line phases, $\theta_j$ ($j=1, \cdots, p$).
The proof is given for $p=1$, but can be generalized to arbitrary $p$.
We assume that every quantity can be regularized in a gauge invariant manner
as in the dimensional regularization method.
Thanks to the invariance under large gauge transformations, $V_\eff(\theta_W)$
is periodic in $\theta_W$ with a period $2\pi$. Thus its Fourier expansion
is written as
\beeq
V_\eff(\theta_W) = \sum_{n=-\infty}^\infty a_n e^{in \theta_W} ~.
\label{effV-Fourier1}
\eneq
The equality is understood as the convergence in the $L^2$ norm.
$a_0$ may be divergent.   The theorem claims that $V_\eff(\theta_W) - a_0$
is finite at the one loop level.

Indeed, the effective action at the one loop level can be written in the 
form 
\beqn
&&\hskip -1cm
V_\eff(\theta_W)^{\rm 1~loop} = 
\sum_j \hat f[ \ell_j \theta_W + c_j ; m_j^2] ~~, \cr
\noalign{\kern 10pt}
&&\hskip -1cm
\hat f(\theta; m^2) = \myfrac{1}{2}  \int \myfrac{d^d q_E}{(2\pi )^d} 
\sum_{n=-\infty}^\infty  
 \ln \bigg\{ q_E^2 + \bigg( n+\myfrac{\theta}{2\pi} \bigg)^2 + m^2 \bigg\} ~~,
\label{proof1}
\eeqn
where $\ell_j$ and $c_j$ are  an integer and a constant, respectively.
$\hat f(\theta; 0) =  f(\theta/2\pi)$ has been explicitly evaluated above
with the  result (\ref{formula6}) which gives a finite contribution to
$V_\eff(\theta_W) - a_0$.  

The argument can be  generalized  for $\hat f(\theta; m^2)$ with 
$m^2 > 0$.   By differentiating  $\hat f$  $d+2$ times with respect to 
$\theta$,  the integral and the infinite sum becomes 
convergent at all $\theta$ for $m^2 > 0$, giving finite 
$\hat f^{(d+2)}$.   By integrating and making use of 
$\int_0^{2\pi} d\theta \, \hat f^{(k)} = 0$ $(k=1, \cdots, d-1)$, 
the finiteness of the $\theta$-dependent part of $\hat f(\theta; m^2)$
is shown.  $\diamondsuit$ 

\vskip .4cm

When $m^2 = 0$, the differentiation of $\hat f(\theta; m^2)$ 
leads to infrared divergence at $\theta=0$ ($mod ~ 2\pi$).  One  
generalizes the theorem.  

\bigskip

\noindent
{\bf Conjecture}
\vskip .2cm 
\centerline{\bigbox{
\vtop{\hsize=14cm
\noindent
The $\theta_W$-dependent part of  $V_\eff(\theta_W)$ is finite almost everywhere 
to every order in perturbation theory.
}}}

\vskip .4cm 
\noindent
{\bf (Outline of proof)} \quad 
There are massless particles whose propagators $D$ take the form
$D^{-1} = p_E^2 + (n + \gamma(\theta_W))^2/R^2$. $D^{-1}$ can vanish
only when $\gamma(\theta_W)$ is an integer.
A point  $\theta_W$ is said to be
regular if $\gamma(\theta_W)$ is not an integer.  

Corrections to $V_\eff(\theta_W)$ at the higher loop levels 
are written as integrals of bubble diagrams.  There are only a finite 
number of diagrams in each order in perturbation theory. 
$\theta_W$ appears in vertices in power, and in propagators $D^{-1}$. 
Hence, by differentiating the diagrams with respect to $\theta_W$ 
at regular points sufficiently
many times, the integrals become convergent. The integrals can diverge
only at a finite number of points in $0 \le \theta_W < 2\pi$.
By integration each diagram gives a finite contribution to the 
$\theta_W$-dependent part of $V_\eff(\theta_W)$ at regular points. 
$\diamondsuit$ 

\vskip .4cm 

\ignore{
It is not clear whether or not $V_\eff(\theta_W) - a_0$ is finite 
at singular points in $\theta_W$. }

\sxn{Dynamical gauge symmetry breaking}
 
Let us consider $SU(N)$ gauge theory on $M^4 \times S^1$ with fermions
in the fundamental and adjoint representations. It can be shown that
all $U$ in (\ref{BC1}) are in one equivalence class of boundary 
conditions, that is, the theory with $\{ U, \beta \}$ is equivalent
with the theory with $\{ I, \beta \}$ on $M^4 \times S^1$.

Without loss of generality we take $U=I$.  Gauge fields are periodic on 
$S^1$. Wilson line phases are related to the zero modes of $A_y$:
\beeq
A_y = \sum_{a=1}^{N^2 - 1} \onehalf A_y^a \lambda^a
= \myfrac{1}{2\pi gR} \pmatrix{\theta_1 \cr &\ddots \cr &&\theta_N\cr} 
\label{Wphase1}
\eneq
where $\sum_{j=1}^N \theta_j = 0$. The four-dimensional effective 
potential is given by
\beqn
&&\hskip -1cm
V_\eff(\theta) = C \Bigg\{
-3 \sum_{j,k=1}^N h_5 \bigg( \myfrac{\theta_j - \theta_k}{2\pi} \bigg) 
+ 4 N_{\rm fund}^F 
\sum_{j=1}^N h_5 \bigg( \myfrac{\theta_j - \beta_{\rm fund}}{2\pi} \bigg) \cr
\noalign{\kern 10pt}
&&\hskip 4cm
+ 4 N_{\rm ad}^F  \sum_{j,k=1}^N h_5 
\bigg( \myfrac{\theta_j - \theta_k- \beta_{\rm ad}}{2\pi} \bigg) 
 ~ \Bigg\} ~~, \cr
\noalign{\kern 10pt}
&&\hskip -.0cm
C = \myfrac{3}{4\pi^2} \myfrac{1}{(2\pi R)^4} ~.
\label{exampleV1}
\eeqn
Here  $N_{\rm fund}^F$ and $N_{\rm ad}^F$ are the numbers of fermion
multiplets in the fundamental and adjoint representations, respectively.
$\beta_{\rm fund}$ and $\beta_{\rm ad}$ are the boundary condition parameters
appearing in (\ref{BC1}). In general, each multiplet of fermions can have
distinct $\beta$. 

\bigskip

\noindent
{\bf Theorem}
\vskip .2cm 
\centerline{\bigbox{
\vtop{\hsize=14cm
\noindent
In pure $SU(N)$ gauge theory on $M^4 \times S^1$, $SU(N)$ gauge 
symmetry is unbroken. 
}}}

\vskip .4cm 
\noindent
{\bf (Proof)} \quad 
This follows immediately from (\ref{exampleV1}) with 
$N^F_{\rm fund} = N^F_{\rm ad}=0$.  $V_\eff$ is minimized when
$\theta_j = \theta_k$ for all $j$ and $k$.  As $\sum_{j=1}^N \theta_j = 0$
($mod ~ 2\pi$), there are $N$ degenerate minima where 
$\theta_j = \theta_k = 0, 2\pi/N, 4\pi/N, \cdots$.  It can be shown that
in pure $SU(N)$ gauge theory on $M^4 \times T^n$, $SU(N)$ gauge 
symmetry is unbroken. 

\vskip .4cm 

\begin{figure}[hbt]
\centering  \leavevmode
\includegraphics[width=6.cm]{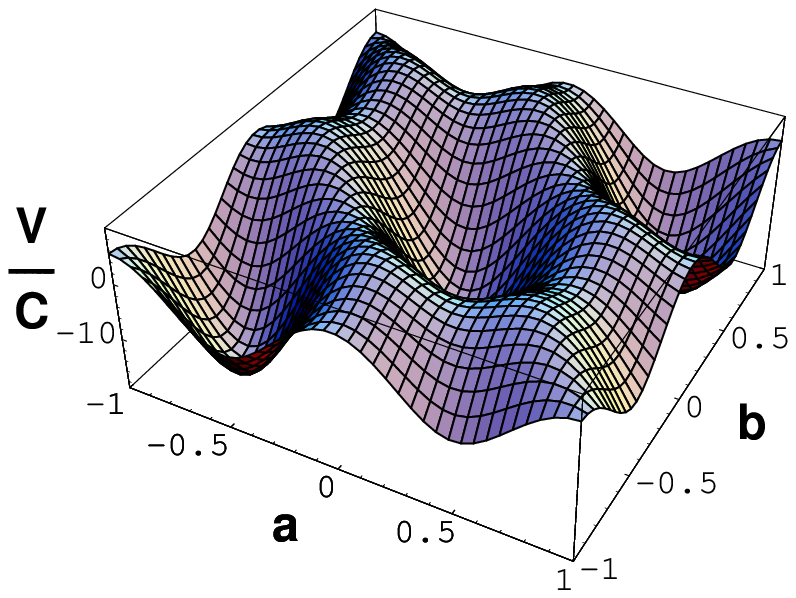}
\hskip 1cm
\includegraphics[width=6.cm]{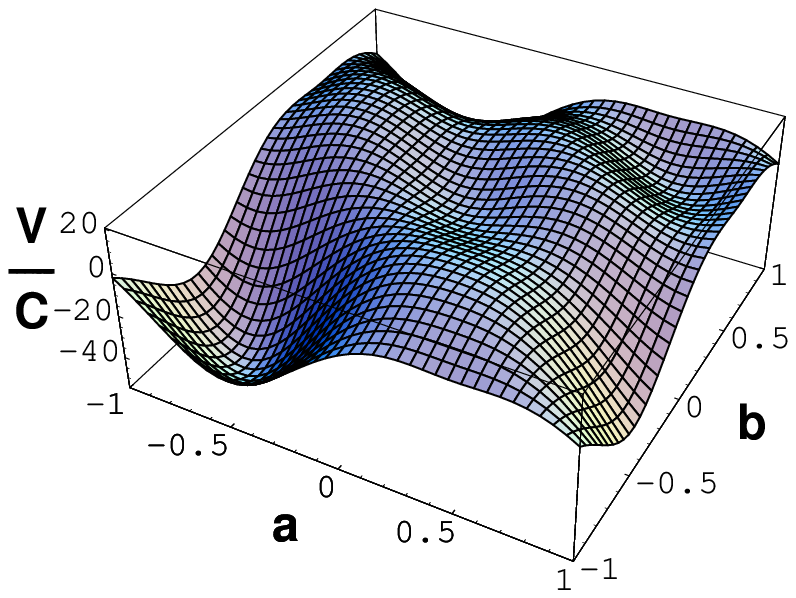}
\vskip -.3cm
\centerline{(a) \hskip 7cm (b)}
\vskip .2cm
\includegraphics[width=6.cm]{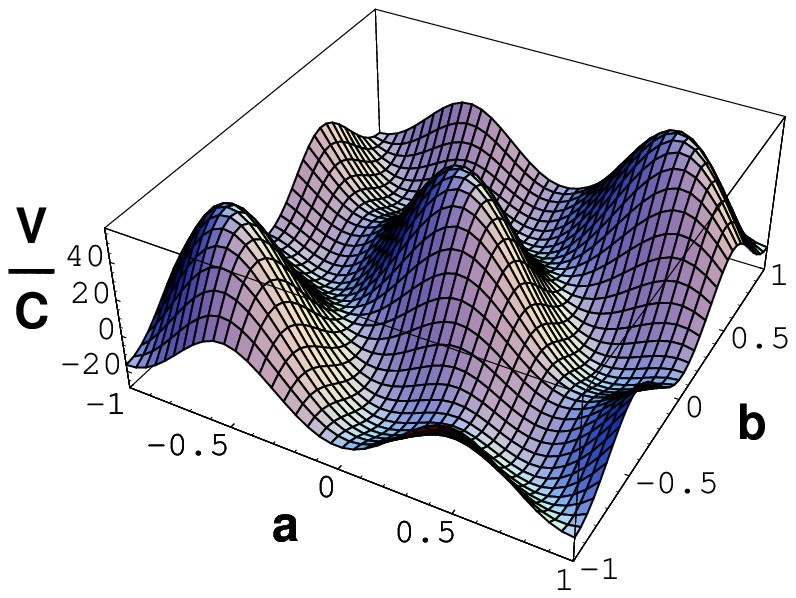}
\hskip 1cm
\includegraphics[width=6.cm]{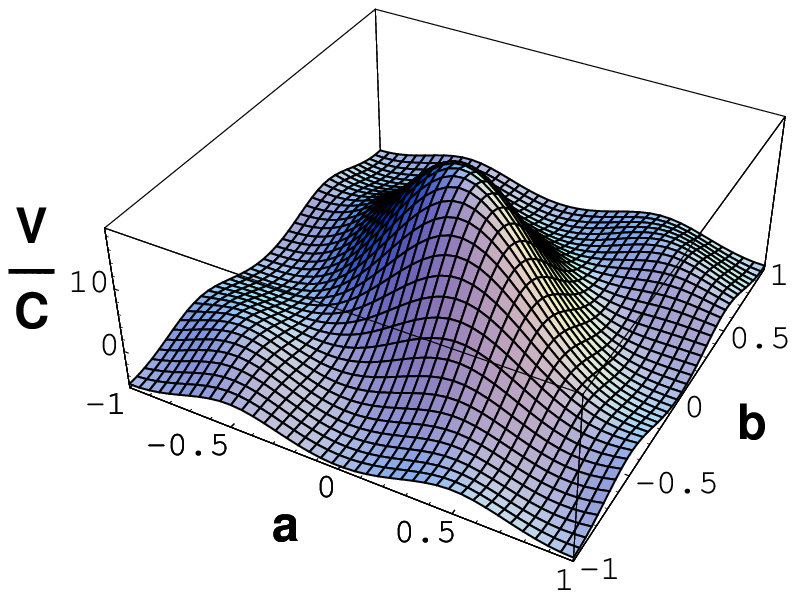}
\vskip -.3cm
\centerline{(c) \hskip 7cm (d)}
\caption{The effective potential $V_\eff (\theta_1, \theta_2)$ in the 
$SU(3)$ gauge theory on $M^4 \times S^1$.  
$(\theta_1, \theta_2) = (\pi a, \pi b)$.
(a) In pure gauge theory.   There are three degenerate minima.
(b) $(N^F_{\rm fund},N^F_{\rm ad})  =(3, 0)$.
$\beta = 0.3 \pi$.  The global minimum is located at 
$(\theta_1,\theta_2) = (- {2\over 3}\pi, - {2\over 3}\pi)$.
(c)  $(N^F_{\rm fund},N^F_{\rm ad})  =(0, 1)$. $\beta = 0$.
The global minimum is located at 
$(\theta_1, \theta_2) = (\pm {2\over 3}\pi, \mp {2\over 3}\pi), 
(0, \pm {2\over 3}\pi), (\pm {2\over 3}\pi, 0)$, which correspond
to $U(1) \times U(1)$ symmetry.
(d) $(N^F_{\rm fund},N^F_{\rm ad})  =(1, 1)$. $\beta = 0$. $\beta = 0.3 \pi$.  
The global minimum is located at 
$(\theta_1, \theta_2) = (\pi, \pi), (0, \pi), (\pi, 0)$, which correspond
to $SU(2) \times U(1)$ symmetry.
}
\label{fig-su3}
\end{figure}

The presence of other matter fields can change the situation.
We list a few examples in $SU(3)$ gauge theory.  In pure gauge 
theory there are three degenerate minima. See fig.\ \ref{fig-su3}(a).
We add fermions to see what happens.

\bigskip
\noindent
(i) $N^F_{\rm fund} > 0$ and $N^F_{\rm ad}=0$

If all fermions are in the fundamental representation and have common
$\beta$,  then the $SU(3)$ symmetry remains unbroken.   The global minimum
of $V_\eff$ is located at 
\beeq
(\theta_1, \theta_2) = \cases{
(- {2\over 3}\pi, -{2\over 3}\pi) &for $0 < \beta < {2\over 3}\pi$,\cr
\noalign{\kern 5pt}
(0,0) &for ${2\over 3}\pi < \beta < {4\over 3}\pi$,\cr
\noalign{\kern 5pt}
( +{2\over 3}\pi, +{2\over 3}\pi) &for ${4\over 3}\pi < \beta < 2\pi$.\cr}
\label{minimum1}
\eneq
See fig.\ \ref{fig-su3}(b).

\bigskip
\noindent
(ii) $N^F_{\rm fund} = 0$ and $N^F_{\rm ad}>0$

Suppose that all fermions belong to the adjoint representation and
have $\beta=0$.  In this case there are six degenerate minima.
The location  $(\theta_1, \theta_2, \theta_3)$ is given by 
a permutation of $(0, {2\over 3} \pi, - {2\over 3} \pi)$.
$SU(3)$ symmetry breaks down to $U(1) \times U(1)$.
See fig.\ \ref{fig-su3}(c).

\bigskip
\noindent
(iii) $N^F_{\rm fund} =N^F_{\rm ad} = 1$

Suppose that there exist fermions in the fundamental representation and 
in the adjoint representation. In particular, consider the case
$N^F_{\rm fund} =N^F_{\rm ad} = 1$ with $\beta=0$.  As depicted in 
fig.\ \ref{fig-su3}(d), the global minima are located at
$(\theta_1, \theta_2, \theta_3) = 
(0, \pi, \pi), (\pi, 0, \pi),(\pi, \pi, 0)$.   The physical
symmetry is $SU(2) \times U(1)$.

\bigskip

We have seen that dynamical gauge symmetry breaking takes place
in the cases (ii) and (iii).  From the four-dimensional viewpoint 
the extra-dimensional components of gauge fields play the role
of Higgs fields in four dimensions.  When Wilson line phases develop
non-trivial vacuum expectation values by quantum effects, gauge
symmetry is spontaneously broken.

Dynamics of Wilson line phases are summarized as follows. 

\bigskip

\centerline{\bigbox{
\vtop{\hsize=14cm
\noindent
{\bf Hosotani mechanism}\\
\qquad 
In gauge theory defined on a non-simply connected  space, 
a configuration with vanishing
field strengths is not necessarily a pure gauge.  There are 
non-Abelian Aharonov-Bohm phases, or Wilson line phases ($\theta_j$), which
become physical degrees of freedom.\\
(i) $\la \theta_j \ra$ is dynamically determined.\\
(ii) When $\la \theta_j \ra$ is nontrivial, gauge symmetry is 
dynamically broken at the quantum level.\\
(iii) Higgs fields in four dimensions are unified with gauge fields.\\
(iv) Physics is the same within each equivalence class of boundary
conditions.
}}}

\bigskip

We add a comment.  In supersymmetric theories the effective potential 
$V_\eff(\theta)$ vanishes if supersymmetry remains exact and unbroken,
as there is cancellation among contributions from bosons and fermions.
When supersymmetry is broken either spontaneously or softly, then 
$V_\eff(\theta)$ becomes nontrivial.  Thus supersymmetry breaking 
can induce dynamical gauge symmetry breaking, as was first clarified
by Takenaga.\cite{Takenaga1}

\vskip 1.5cm

\leftline{\bf \large Part II.  \vtop{\hsize=12cm \noindent
Dynamical gauge symmetry breaking in the\\  electroweak theory}}

\sxn{Electroweak gauge-Higgs unification on orbifolds}

There are two important ingredients to be implemented in applying
the scheme of dynamical gauge-Higgs unification to the electroweak 
interactions.  First of all the electroweak symmetry is
$SU(2)_L \times U(1)_Y$, which is broken to $U(1)_{EM}$.  
In the standard electroweak theory the Higgs field in an $SU(2)_L$ 
doublet induces the symmetry breaking.   In the scheme of dynamical
gauge-Higgs unification explained in the Part I, Higgs fields in four
dimensions are identified with the extra-dimensional components of
gauge fields which necessarily belong to the adjoint representation
of the gauge group.  Thus the Higgs doublet in the electroweak theory must
be a part of the field in the adjoint representation of a larger
group, as was  first clarified by Fairlie\cite{Fairlie1} 
and by Forgacs and Manton.\cite{Manton1}
The enlarged gauge group has to contain either $SU(3)$, $SO(5)$, or 
$G_2$.

Secondly fermions are chiral in the electroweak theory.  The most
economical and powerful way of having chiral fermions in four 
dimensions is to start a gauge theory on orbifolds.\cite{Pomarol1}  
The orbifold projection makes the fermion content chiral.

Many models have been proposed in the 
literature.\cite{Pomarol1}-\cite{HHKY}  As an 
extra-dimensional space, $S^1/Z_2$ and $T^2/Z_2$ have been
most commonly considered.  Gauge theory on the Randall-Sundrum
warped spacetime has been intensively investigated, as well.

Let us consider $SU(N)$ gauge theory on $M^4 \times (T^n/Z_2)$
with coordinates $x^\mu$ ($\mu=0, \cdots, 3$) and 
$y^a$ $(a=1, \cdots, n)$.\cite{HNT1}  
$T^n/Z_2$ is obtained by the identification
\beqn
T_a &:&  \vec{y}+\vec{l}_a ~\sim~ \vec{y} \cr
\noalign{\kern 5pt}
&& \vec{l}_a = (0, \cdots, 2\pi R_a, \cdots, 0) 
  ~~~(a=1, 2, \cdots, n)~, \cr
\noalign{\kern 5pt}
Z_2 &:&  - \vec{y} ~\sim ~\vec{y} ~~~.
\label{orbifold1}
\eeqn
As $T^n$ is not simply connected, there appear Wilson line phases
as physical degrees of freedom as explained in Part I. 
In the course of the $Z_2$ orbifolding there appear fixed points.
Theory requires additional boundary
conditions at those fixed points, which gives us benefit of eliminating
some of light modes in various fields.  Chiral fermions naturally
appear at low energies.  Some of Wilson line phases drop out from
the spectrum, while the others survive.  The surviving Wilson line
phases play the role of Higgs fields in $M^4$, inducing dynamical
gauge symmetry breaking.

Although $(x, \vec y)$ and $(x, \vec y + \vec l_a)$ represent the same
point on $T^n$, the values of fields need not be the same. In general 
\beqn
&&\hskip -1cm 
A_M(x, \vec y + \vec l_a) =
U_a A_M(x, \vec y ) \, U_a^\dagger  ~~~, \cr
\noalign{\kern 5pt}
&&\hskip -1cm 
\psi(x, \vec y + \vec l_a) 
  =  \eta_a \, T[U_a] \psi(x, \vec y) ~~~, \cr
\noalign{\kern 5pt}
&&\hskip -1cm 
[U_a , U_b ] = 0 ~~~, ~~~ U_a \in SU(N) ~~~(a,b =1,\cdots, n)~.
\label{BC5}
\eeqn
$\eta_a$ is a $U(1)$ phase factor.  $T[U_a] \psi= U_a \psi$ or   
$U_a \psi U_a^\dagger$ for $\psi$ in the fundamental or adjoint representation,
respectively. The boundary condition (\ref{BC5}) guarantees that the physics is 
the same at $(x, \vec y)$ and $(x, \vec y + \vec l_a)$.  The condition
$[U_a , U_b ] = 0$ is necessary to ensure $T_a T_b = T_b T_a$.

Similar conditions follow from the $Z_2$ orbifolding:
$Z_2 ~:~  - \vec{y} ~\sim ~\vec{y}$.   
On $T^n$, this parity operation allows fixed points at $z$ where
the relation $\vec{z}=-\vec{z}+\sum_a m_a\vec{l}_a$ ($m_a=$ an integer)
is satisfied.  There appear $2^n$ fixed points on $T^n$. Combining it  with
loop translations in (\ref{BC5}),  one finds that parity
around each fixed point is also  symmetry:
\beeq
Z_{2, j} ~:~  \vec z_j - \vec{y} ~\sim ~ \vec z_j + \vec{y} 
  \quad (j=0, \cdots, 2^n -1) ~~~.
\label{parity2}
\eneq
Accordingly fields must satisfy additional boundary conditions.  

Let spacetime be $M^4 \times (T^2/Z_2)$, in which case
$\vec z_0 =(0,0)$, $\vec z_1 =(\pi R_1,0)$, $\vec z_2 =(0,\pi R_2)$,
and $\vec z_3 =(\pi R_1, \pi R_2)$. 
Under $Z_{2,j}$ in (\ref{parity2})
\beqn
&&\hskip -1cm 
\pmatrix{A_\mu \cr A_{y_a} \cr} (x, \vec z_j - \vec y) =
P_j \pmatrix{A_\mu \cr - A_{y_a} \cr} (x, \vec z_j + \vec y) \, 
   P_i^\dagger ~ ~, \cr
\noalign{\kern 10pt}
&&\hskip -1cm 
\psi(x, \vec z_j - \vec y) = \eta_j' \, T[P_j] \, (i\Gamma^4\Gamma^5)
\psi(x, \vec z_j + \vec y) \qquad (\eta_j' = \pm 1) \cr
\noalign{\kern 5pt}
&&\hskip -1cm 
\qquad (a=1,2, \quad j=0, 1, 2,3) ~~.
\label{BC6}
\eeqn
Here $P_j = P_j^{-1} = P_j^\dagger \in SU(N)$.  Not all $U_a$'s and $P_j$'s
are independent.  On $T^2/Z_2$, only three of them are independent.  One can show
that
\beqn
&&\hskip -1cm 
U_a = P_a P_0 ~~~,~~~
P_3 = P_2 P_0 P_1 = P_1 P_0 P_2 ~~~, \cr
\noalign{\kern 5pt}
&&\hskip -1cm 
\eta_a = \eta_0' \eta_a' = \pm 1  \quad (a=1,2) ~. 
\label{BC7}
\eeqn 
Gauge theory on $M^4 \times (T^2/Z_2)$ is specified with a set of
boundary conditions $\{ P_j , \eta_j' ~;~ j=0, 1, 2 \}$. 
If fermions $\psi$ in (\ref{BC6}) are 6-D Weyl fermions, i.e. 
$\Gamma^7 \psi = +\psi$ or $-\psi$ where $\Gamma^7 = \Gamma^0 \cdots \Gamma^5$,
then the boundary condition (\ref{BC6}) makes 4D fermions chiral.

At a first look, the original gauge symmetry is broken by the boundary
conditions if $P_0$, $P_1$ and $P_2$ are not proportional to the identity
matrix. This part of the symmetry breaking is often called the orbifold
symmetry breaking in the literature.  However, the physical
symmetry of the theory can be different from the symmetry of the boundary 
conditions, and different sets of boundary conditions can be equivalent
to each other.

\sxn{The Hosotani mechanism on orbifolds}

It is important to recognize that sets of boundary conditions form
equivalence classes.  As in (\ref{BC2}), 
under a gauge transformation (\ref{gaugeT1})
$A_M'$ obeys a new set of boundary conditions $\{ P_j' , U_a' \}$ where
\beqn
&&\hskip -1cm
P_j' =\Omega(x, \vec z_j -\vec y) \, P_j 
   \,\Omega(x, \vec z_j +\vec y)^\dagger ~, \cr
\noalign{\kern 5pt}
&&\hskip -1cm
U_a' = \Omega(x, \vec y + \vec l_a)\, U_a\, \Omega(x, \vec y)^\dagger ~, \cr
\noalign{\kern 5pt}
&&\hskip 0. cm
\hbox{provided ~} \dd_M P_j' = \dd_M U_a' = 0 ~~.
\label{newBC1}
\eeqn 
The set $\{ P_j' \}$ can be different from the set $\{ P_j \}$.  
When the relations in (\ref{newBC1}) are satisfied, we write
\beeq
\{ P_j' \} ~\sim~ \{ P_j \} ~~.
\label{relation11}
\eneq
This relation is transitive, and therefore is an equivalence
relation.  Sets of boundary conditions form equivalence classes of boundary 
conditions with respect to the equivalence 
relation (\ref{relation1}). \cite{YH2, HHHK,  HHK}

The equivalence relation (\ref{relation1}) indeed implies the equivalence of
physics as a result of dynamics of Wilson line phases.  Wilson line phases
are zero modes ($x$- and $\vec y$-independent modes) 
of extra-dimensional components of gauge fields which satisfy
\beqn
&&\hskip -1cm
A_{y_a} = \sum_{\alpha \in H_W} \onehalf A_{y_a}^\alpha
     \lambda^\alpha ~~~, ~~~
     [A_{y_a} , A_{y_b} ] = 0 ~~~,  ~~~ (a,b=1, \cdots, n) ~, \cr
\noalign{\kern 5pt}
&&\hskip -1cm
H_W = \Big\{ ~ \lambda^\alpha ~ ; 
 ~ \{ \lambda^\alpha , P_j \} = 0 \quad (j= 0, \cdots, 2^n - 1) ~ \Big\} ~.
\label{wilson1}
\eeqn
Consistency with the boundary condition (\ref{BC3}) requires 
$\lambda^\alpha$ in the sum to belong to $H_W$.  
Given the boundary conditions, these Wilson line phases cannot be 
gauged away.  They are physical degrees of freedom.  They label
degenerate classical vacua,
parametrizing flat directions in the classical potential.
The values of $\la A_{y_a} \ra$ are determined, at the quantum level, 
 from the location of the 
absolute minimum of  the effective potential $V_\eff [ A_{y_a} ]$.

Other than the restriction to $\lambda^a$ in (\ref{wilson1}), the 
situation is the same as in gauge theory on $M^4 \times S^1$ discussed
in Part I. 
Physical symmetry is determined in the combination of the 
boundary conditions $\{ P_j , \eta_j' \}$ and the expectation values of
the Wilson line phases $\la A_{y_a} \ra$.   Physical symmetry is, 
in general, different from the symmetry of the boundary conditions.
As a result of quantum dynamics gauge symmetry can be dynamically broken
by Wilson line phases.
This is called the Hosotani mechanism.   The summary given at the end of
Section 4 remains valid in gauge theory on orbifolds as well.

In gauge theory on $M^4 \times T^n$, there is only one equivalence
class of boundary conditions.  On $M^4 \times (T^n/Z_2)$, however, 
there are more than one equivalence classes.  In $SU(N)$ gauge 
theory on $M^4 \times (S^1/Z_2)$, for instance, there are $(N+1)^2$
equivalence classes.\cite{HHK}  If one of the $P_j$'s is proportional
to the identity matrix, then there is no $\lambda^a$ belonging to
$H_W$ in (\ref{wilson1}) so that there is no Wilson line phase.
The fact that there are multiple equivalence classes of boundary conditions
gives rise to the arbitrariness problem of boundary conditions.\cite{YH5}
It is desirable to show how and why one particular equivalence class of
boundary conditions is selected by dynamics.

\sxn{$U(3)_S \times U(3)_W$ model on $M^4 \times (T^2/Z_2)$}

To achieve dynamical gauge-Higgs unification in the electroweak theory 
one has to enlarge the gauge group such that doublet Higgs fields in 
$SU(2)_L$ can be identified with a part of gauge fields in the enlarged 
group $\hat G$.  
The original proposal by Manton was along this line, but the
resultant low energy theory was far from the reality.
Antoniadis, Benakli and Quiros proposed an intriguing model in which 
$\hat G$ is taken to be $U(3)_S \times U(3)_W$ with gauge couplings 
$g_S$ and $g_W$.\cite{gaugeHiggs1}
$U(3)_S$ is ``strong'' $U(3)$ which decomposes to 
color $SU(3)_c$ and $U(1)_3$.  $U(3)_W$ is ``weak'' $U(3)$ which decomposes 
to  weak $SU(3)_W$ and $U(1)_2$.  The theory
is defined on $M^4 \times (T^2/Z_2)$.  Boundary conditions at
fixed points of $T^2/Z_2$ are chosen to be
\beeq
P_0 = P_1 = P_2 = 
I_S \otimes \pmatrix{-1 \cr &-1\cr &&+1\cr}_W ~~.
\label{BC8}
\eneq
The boundary condition (\ref{BC8}) breaks $SU(3)_W$ to
$SU(2)_L \times U(1)_1$ at the classical level.  
There are three $U(1)$'s left over.

Fermions obey boundary condition described in (\ref{BC6}).  Let 
$(n_S, n_W)^\sigma$ stand for a fermion in the $n_S$ ($n_W$)
representation of $U(3)_S$ ($U(3)_W$)  with 6D-Weyl
eigenvalue $\Gamma^7 = \sigma$.   Three generations of
leptons are assigned as follows.  Leptons are
\beeq
L_{1,2,3} = (1 , 3)^+ ~~~:~~~
\pmatrix{{\nu_L} \cr {e_L} \cr \tilde e_L \cr} ~,~
\pmatrix{\tilde \nu_R \cr \tilde e_R \cr  {e_R} \cr}~~ \hbox{etc}.
\label{lepton1}
\eneq
Similarly, for right-handed down quarks we have
\beeq
D^c_{1,2,3} = (\bar 3 , 1)^+ ~~~:~~~
{d^c_L}  ~~,~~ {\tilde d^c_R} ~~ \hbox{etc}.
\label{quark1}
\eneq
For other quarks, each generation has its own distinct assignment:
\beqn
Q_{1}  = (3 , \bar 3)^+ 
&&
\pmatrix{{u_L} \cr {d_L} \cr \tilde u_L \cr} ~,~
\pmatrix{\tilde u_R \cr \tilde d_R \cr  {u_R} \cr} \cr
\noalign{\kern 10pt}
Q_{2}  = (3 , \bar 3)^-
&&
\pmatrix{{s_L} \cr {c_L} \cr \tilde c_L \cr} ~,~
\pmatrix{\tilde s_R \cr \tilde c_R \cr  {c_R} \cr} \cr
\noalign{\kern 10pt}
Q_{3}  = (\bar 3 ,  3)^-
&&
\pmatrix{\tilde b_L^c \cr \tilde t_L^c \cr {t_L^c} \cr} ~,~
\pmatrix{{b_R^c} \cr {t_R^c} \cr \tilde t_R^c \cr} ~.
\label{quark2}
\eeqn
Due to the boundary conditions either $SU(2)_L$ doublet
part or singlet part has zero modes.  In (\ref{lepton1})-(\ref{quark2}), 
fields with tilde $\tilde {~}$ do not have zero modes.

With these assignments of fermions only one combination of
three $U(1)$ gauge groups remains anomaly free, which is
identified with weak hypercharge $U(1)_Y$.  Gauge bosons 
corresponding to the other two combinations of three $U(1)$ 
gauge groups become massive by the Green-Schwarz mechanism.
Hence, the remaining symmetry at this level is
$SU(3)_C \times SU(2)_L \times U(1)_Y$.

\def\myb{{\vphantom{\myfrac{1}{2}}}}

The metric of $T^2$ is given by
\beeq
ds^2 = dy_1^2 + 2 \cos \theta dy_1 dy_2 + dy_2^2 ~,
\label{metric1}
\eneq
where $\theta$ is the angle between the directions of the
$y_1$ and $y_2$ axes.  
There are Wilson line phases in the $SU(3)_W$ group.  They 
are 
\beeq
A_{y_j} 
={1\over \sqrt{2}} \pmatrix{~ & \myb \Phi_j \cr
         ~~\Phi_j^\dagger ~ &  \cr}
         \quad (j=1,2) ~.
\label{wilson4}
\eneq
$\Phi_1$ and $\Phi_2$ are $SU(2)_L$ doublets.  The resultant
theory is the Weinberg-Salam theory with two Higgs doublets.
The classical potential for the Higgs fields results from
the $(F_{y_1 y_2})^2$ part of the gauge field action:
\beeq
V_{\rm tree}(\Phi_1, \Phi_2) =  \myfrac{g_W^2}{2 \sin^2 \theta} 
\Big\{ 
 \Phi_1^\dagger \Phi_1^{} \cdot \Phi_2^\dagger \Phi_2^{} 
+ \Phi_2^\dagger \Phi_1^{} \cdot \Phi_1^\dagger \Phi_2^{}
-(\Phi_2^\dagger \Phi_1^{})^2 - (\Phi_1^\dagger \Phi_2^{})^2 
\Big\} ~~.
\label{tree1}
\eneq
There is no quadratic term.  The potential (\ref{tree1}) is
positive definite and  has flat directions.  The potential
vanishes if $\Phi_1$ and $\Phi_2$ are proportional to each other
with a real proportionality constant.

To determine if the electroweak symmetry is dynamically broken, 
one need to evaluate quantum corrections to the effective potential
of $\Phi_1$ and $\Phi_2$.\cite{HNT2}    The effective potential in the flat 
 directions is obtained, without loss of generality, for 
 the configuration 
 \beeq
 2 g_W R_1 A_{y_1} = \pmatrix{&&0\cr &&a\cr 0&a\cr} ~~,~~
 2 g_W R_2 A_{y_2} = \pmatrix{&&0\cr &&b\cr 0&b\cr} ~~,
 \label{wilson5}
 \eneq
where $a$ and $b$ are real.
Our task is  to find $V_\eff(a,b)$ and thereby determine 
the physical vacuum. 

Depending on the location of the global minimum of $V_\eff(a,b)$,
the physical symmetry varies.  It is given by
\beeq
(a,b) = \cases{
(0,0)  &$\Rightarrow$ ~ $SU(2)_L \times U(1)_Y$\cr
\noalign{\kern 5pt}
(0,1), (1,0), (1,1) ~~~&$\Rightarrow$ ~ $U(1)_{EM} \times U(1)_Z$\cr
\noalign{\kern 5pt}
\hbox{otherwise} &$\Rightarrow$ ~ $U(1)_{EM}$.\cr}
\label{symmetry5}
\eneq
For generic values of $(a,b)$, electroweak symmetry breaking
takes place.  The Weinberg angle is given by
\beeq
\sin^2 \theta_W
= {1\over 4 + \myfrac{2 g_W^2}{3 g_S^2}} ~~,
\label{Wangle}
\eneq 
which can be close to the observed value.  A small deviation 
from the value $0.25$ is brought by  $g_W^2/g_S^2$.
We note that
in the $SU(3)_c \times SU(3)_W$ model the Weinberg angle
turns out too large.

The evaluation of $V_\eff(a,b)$ is straightforward.  In the 
non-supersymmetric model the matter content is given by
gauge fields (including ghosts) and fermions summarized in
(\ref{lepton1})-(\ref{quark2}).  Only gauge fields in $SU(3)_W$
give ($a,b$)-dependent contributions.  The result is
\beeq
V_\eff(a,b)  
= (8 - 16 N_F) \cdot I\Big( {a\over 2}, {b\over 2} \Big) 
    + 4 \cdot I(a,b) + \hbox{const.} 
\label{Veff2}
\eneq
where
\beqn
&&\hskip -1cm
I(a,b) = - {1\over 16\pi^9} \Bigg\{ ~
\sum_{n=1}^\infty {\cos 2\pi n a \over n^6 R_1^6}
 +   \sum_{m=1}^\infty {\cos 2\pi m b \over m^6 R_2^6} \cr
\noalign{\kern 10pt}
&&\hskip 2.cm 
 + \sum_{n=1}^\infty  \sum_{m=1}^\infty 
 {\cos 2\pi (n a + m b) \over (n^2 R_1^2 + m^2 R_2^2 + 2nm R_1 R_2 \cos\theta)^3}
 \cr
\noalign{\kern 10pt}
&&\hskip 2.cm 
+ \sum_{n=1}^\infty  \sum_{m=1}^\infty 
 {\cos 2\pi (n a - m b) \over (n^2 R_1^2 + m^2 R_2^2 - 2nm R_1 R_2 \cos\theta)^3}
~ \Bigg\} ~~.
\label{Veff3}
\eeqn
$N_F=3$ in the minimal model.

\begin{figure}[bt]
\centering  \leavevmode
\includegraphics[width=6.cm]{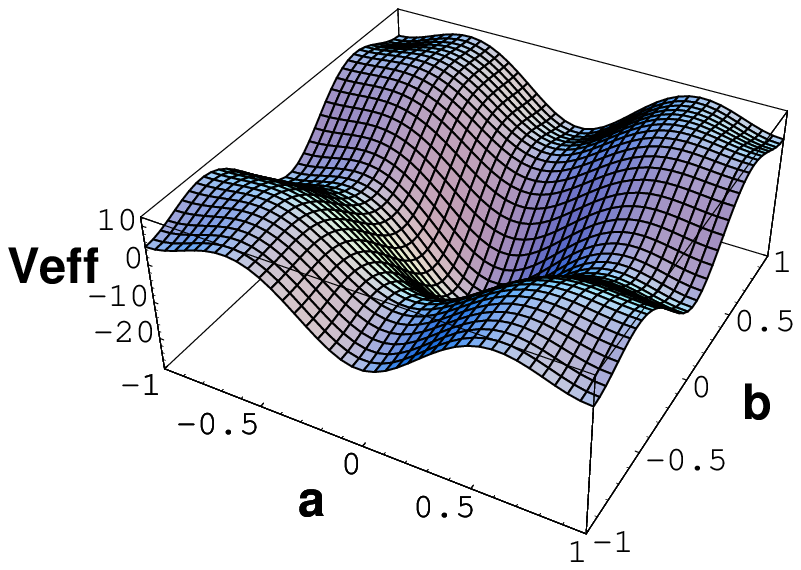}
\hskip 1cm
\includegraphics[width=6.cm]{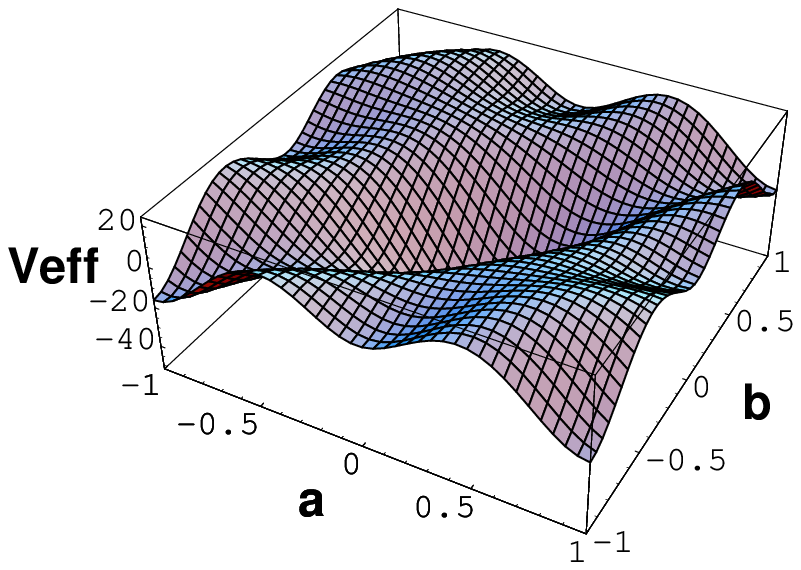}
\vskip -.3cm
\centerline{(a) \hskip 7cm (b)}
\vskip .2cm
\includegraphics[width=6.cm]{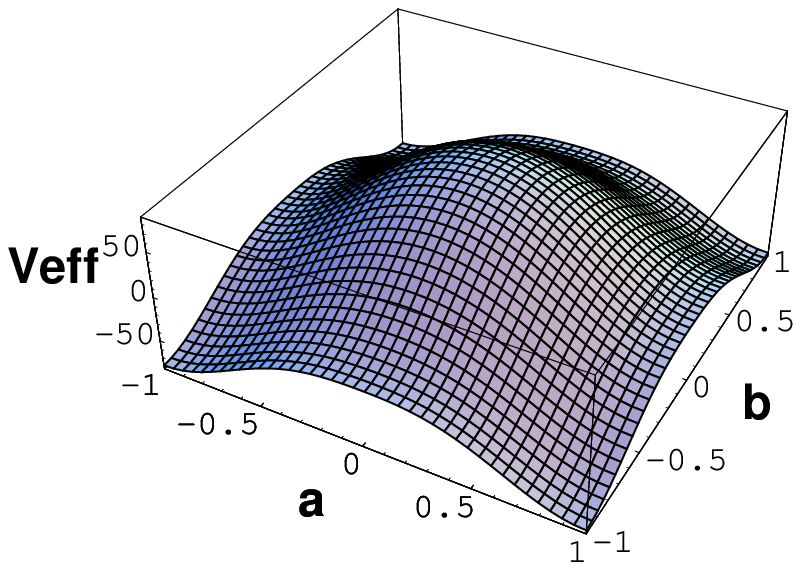}
\hskip 1cm
\includegraphics[width=6.cm]{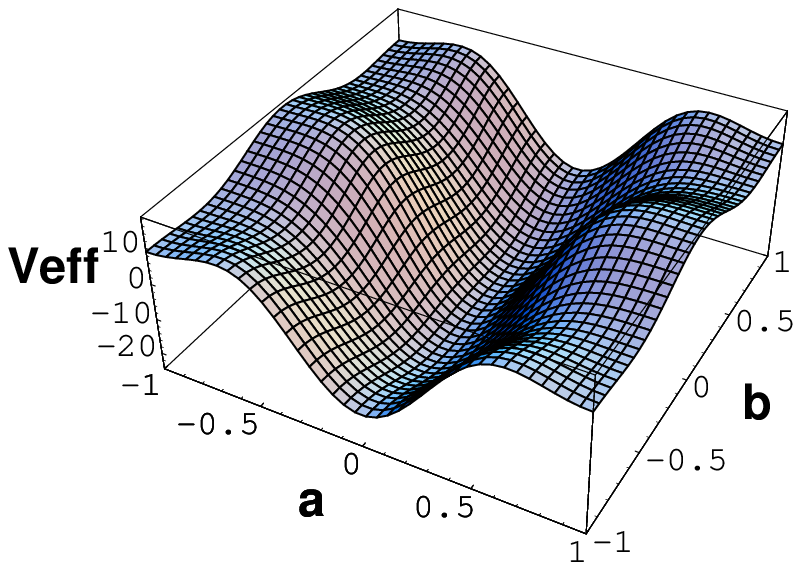}
\vskip -.3cm
\centerline{(c) \hskip 7cm (d)}
\caption{The effective potential $V_\eff (a,b)$ in the 
$U(3) \times U(3)$ gauge theory on $M^4 \times (T^2/Z_2)$ with $R_1=R_2$.  
(a) In pure gauge theory with $\cos \theta = 0$.  
(b) In pure gauge theory with $\cos \theta = 0.6$.
(c) In the presence of a minimal set of fermions with  $\cos \theta = 0$.
The global minimum is located at 
$(a, b) = (1,1)$   which corresponds
to $U(1)_{EM} \times U(1)_Z$ symmetry.
(d) In the presence of parity partners of quarks and leptons and one fermion
in the adjoint representation with $\cos \theta = 0$.
The minimum is located at  $(a,b) = (0, \pm 0.269)$.
The electroweak symmetry breaks down to $U(1)_{EM}$.
}
\label{fig-u3u3}
\end{figure}

If there were no fermions, i.e., $N_F=0$, $V_\eff(a,b)$ has the global 
minimum at $(a,b) = (0,0)$ for any value of $\cos \theta$ 
so that $SU(2)_L \times U(1)_Y$ symmetry is unbroken.  See fig.\ 
\ref{fig-u3u3} (a) and (b).

In the presence of fermions, the point $(a,b) = (0,0)$
becomes unstable.  The global minimum is located at
$(a,b) = (1,1)$ for $|\cos\theta| < 0.5$ and 
at $(a,b) = (0,1)$ or $(1,0)$ for $|\cos\theta| >0.5$.
In either case the residual symmetry is $U(1)_{EM} \times U(1)_Z$.
Although the $SU(2)_L$ symmetry is partially
broken and $W$ bosons acquire masses, $Z$ bosons remain massless.
See fig.\ \ref{fig-u3u3}(c).

Models having the electroweak symmetry breaking are obtained 
by adding heavy fermions.  For each quark/lepton multiplet in 
(\ref{lepton1})-(\ref{quark2}), 
which has $(\eta_1,  \eta_2) = (1,1)$
in (\ref{BC5}),  we introduce three parity partners with
$(\eta_1, \eta_2) = (-1,1), (1,-1), (-1,-1)$. Further we
add fermions in the adjoint representation with 
$( \eta_1, \eta_2) = (-1,1)$. The total effective potential is,
up to a constant, 
\beqn
&&\hskip -1cm
V_\eff(a,b)^{\rm total} = 
8 I(\onehalf a, \onehalf b)  + 4 I(a,b) 
- N_{Ad} \Big\{ 8 I(\onehalf a + \onehalf, \onehalf b)  
   + 4 I(a + \onehalf,b) \Big\} \cr
\noalign{\kern 10pt}
&&\hskip -0.5cm
 - 16 N_F \Big\{ I(\onehalf a, \onehalf b )  
 + I(\onehalf a + \onehalf, \onehalf b)
 + I(\onehalf a , \onehalf b + \onehalf)
 + I(\onehalf a + \onehalf, \onehalf b + \onehalf) \Big\} ~.
\label{Veff4}
\eeqn
Here $N_{Ad}$ and $N_F$ are the numbers of Weyl fermions in the adjoint
representation and of generation of quarks and leptons, respectively.
Fermions with $(\eta_1,  \eta_2) \not= (1,1)$ do not have
zero modes.  For $N_F=3$ the spectrum at low energies is the same as 
in the minimal model.

With $N_F=3$, $N_{Ad}=1$ and $R_1=R_2$ in (\ref{Veff4}), the global minima 
of $V_\eff(a,b)^{\rm total}$ are located at $(a,b)=(0, \pm 0.269)$
for $\cos\theta=0$ and at at $(a,b)=(\pm 0.013, \pm 0.224)$ for
$\cos\theta = 0.1$.  For $\cos\theta > 0.133$ the global minimum
is located at $(a,b)=(0,0)$.  The electroweak
symmetry is dynamically broken for $|\cos\theta | < 0.133$.  

\sxn{$m_W$, $m_H$ and $M_{KK}$}

One of the most intriguing features in the dynamical gauge-Higgs 
unification is that the mass of the Higgs boson, $m_H$,  and the 
energy scale of the Kaluza-Klein excitations, $M_{KK}$,  are predicted 
in terms of the W boson mass, $m_W$.  Wilson line phases play the 
role of Higgs fields in four dimensions.  $m_W$ is determined
from the  Wilson line phases and the size of extra dimensions, whereas
$m_H$ is determined from the effective potential for the Wilson 
line phases.\cite{HNT2}

The W boson mass arises from the $\tr F_{\mu y_j} F^{\mu y_j}$ 
term in the Lagrangian.  Non-vanishing $(a,b)$ gives
\beeq
m_W^2 =
\frac{1}{4 \sin^2\theta} 
\left( \frac{a^2}{R_1^2} +\frac{b^2}{R_2^2} 
- \frac{2a b \cos\theta }{R_1 R_2} \right)  ~. 
\label{spectrum1}
\eneq
In the model described in (\ref{Veff4}) with 
$N_F=3$, $N_{Ad}=1$ and $R_1=R_2=R$, one finds that
\beeq
m_W = \cases{
\myfrac{0.135}{R} &for $\cos\theta = 0$,\cr
\myfrac{0.112}{R} &for $\cos\theta = 0.1$.\cr}
\label{spectrum2}
\eneq

On $M^4 \times (T^2/Z_2)$ there appear two Higgs doublets in four 
dimensions.  Three of the eight degrees of freedom are
absorbed by $W$ and $Z$ bosons.  A charged Higgs particle acquires
a mass $\sim m_W$, while a neutral CP-odd Higgs particle  acquires
a mass $\sim 2 m_W$.  The most problematic is the mass of two neutral
CP-even Higgs particles, which correspond to quantum fluctuations
in the directions of the Wilson line phases.  By making use of 
(\ref{wilson4}) and (\ref{wilson5}), the masses are evaluated from
the two eigenvalues of the matrix 
$K^{jk} = g^2 R^2 (\dd^2 V_\eff / \dd a_j \dd a_k)|_{\rm min}$
where $(a_1, a_2) = (a,b)$.  They are given by
\beeq
m_H = \left\{ \matrix{ (0.871, 3.26) \cr (0.799, 4.01) } \right\}
\times \bigg( \myfrac{g_4^2}{4\pi} \bigg)^{1/2} ~ m_W 
\qquad {\rm for ~} \cos\theta = \cases{0 \cr 0.1 \cr}
\label{Higgs1}
\eneq
where the four-dimensional gauge coupling constant is related to
the six-dimensional one by  $g_4^2 = g^2/ (2\pi^2 R^2 \sin\theta)$.  

(\ref{spectrum2}) and (\ref{Higgs1}) show how $M_{KK} = 1/R$
and $m_H$ are related to $m_W$ in this scheme.  From (\ref{spectrum2}) 
$M_{KK}$ turns out $(7 \sim 9) m_W$, being too low from the viewpoint 
of the observational limit.  As inferred from (\ref{spectrum1}), 
$m_W$ is given, generically in flat space, by
\beeq
M_{KK} \sim 2 ~ \myfrac{\pi}{\theta_W} ~ m_W ~~.
\label{KKscale1}
\eneq
As the minimum of the effective potential for the Wilson line phase
$\theta_W$ is located typically at $\theta_W = (0.2 \sim 0.4) \pi$,
$M_{KK}$ appears in the range $400 \sim 800 \,$GeV.   
Further, it follows from (\ref{Higgs1}) that the  mass of 
the lightest Higgs particle is about 10$\,$GeV, which contradicts 
with the observation.  In general one finds, in flat space, 
\beeq
m_H \sim 0.2 ~ \sqrt{\alpha_w} 
~ \myfrac{\pi}{\theta_W} ~ m_W  ~~, 
\label{Higgs2}
\eneq
where $\alpha_w = {g_4^2}/{4\pi}$.  As the Higgs mass is generated
by radiative corrections, there appears 
the factor $\sqrt{\alpha_w}$ which leads to a small Higgs mass.

\sxn{Prospect}

We have shown  that the dynamical gauge-Higgs unification 
is achieved in higher dimensional gauge theory.  Higgs fields are
identified with Wilson line phases in gauge theory.  Dynamical
symmetry breaking is induced by the Hosotani mechanism.

In the dynamical gauge-Higgs unification $M_{KK}$ and $m_H$ are
related to $m_W$ and $\theta_W$.  The results (\ref{KKscale1}) and
(\ref{Higgs2}) generically follow when the extra-dimensional space
is flat.   With a typical value $\theta_W = (0.2 \sim 0.4) \pi$,
both $M_{KK}$ and $m_H$ turn out too low.

How can we circumvent this difficulty?  One way is to have a model
in which the global minimum of the effective potential is located
at very small $\theta_W$.  This goal is, in principle, achieved by
tuning the matter content.  However, it  usually requires 
to incorporate many additional fields so that the resultant 
theory is not realistic.

More promising is to consider a gauge theory in higher dimensions
where the extra-dimensional space is curved.  Gauge theory defined 
in the Randall-Sundrum warped spacetime is particularly 
interesting.\cite{RS}-\cite{Mabe}
It has been recently shown \cite{Mabe} that  dynamical gauge-Higgs 
unification in the Randall-Sundrum warped spacetime leads to,
in place of (\ref{KKscale1}) and (\ref{Higgs2}),
\beqn
&&\hskip -1cm
M_{KK} \sim \sqrt{2\pi kR} ~ \myfrac{\pi}{\theta_W} ~ m_W ~~, \cr
\noalign{\kern 5pt}
&&\hskip -1cm
m_H \sim 0.3 ~ kR ~ \sqrt{\alpha_w} 
~ \myfrac{\pi}{\theta_W} ~ m_W  ~~.
\label{RS1}
\eeqn
Here $k^2$ and $R$ are the curvature and size of the extra-dimensional
space.  If  the structure of spacetime is 
determined at the Planck mass scale, then  $k = O(M_{Pl})$.
To have $m_W \sim 80\,$GeV, the relation $k = O(M_{Pl})$
in turn leads to  $kR = 12 \pm 1$.  $kR$ is not a free 
parameter in the dynamical gauge-Higgs unification scheme.
With a typical value $\theta_W = (0.2 \sim 0.4) \pi$,
it is predicted that $M_{KK} = (1.7 \sim 3.5) \,$TeV and 
$m_H = (140 \sim 280) \,$GeV.

It is very exciting that the mass of the Higgs particle is 
predicted in the region where the experiments at LHC can
certainly explore.  Detailed analysis of the interactions of
the Higgs particles in the dynamical gauge-Higgs unification
scheme will shed light on what to explore in the LHC experiments.
We might be able to observe the dynamical gauge symmetry 
breaking by the Wilson line phases.

\vskip .5cm

\leftline{\bf Acknowledgment}
This work was supported in part by  Scientific Grants from the Ministry of 
Education and Science, Grant No.\ 13135215, 
Grant No.\ 15340078, Grant No.\ 17043007, and Grant No.\ 17540257.


\def\jnl#1#2#3#4{{#1}{\bf #2} (#4) #3}

\def\Zphys{{\em Z.\ Phys.} }
\def\jssc{{\em J.\ Solid State Chem.\ }}
\def\jpsJ{{\em J.\ Phys.\ Soc.\ Japan }}
\def\ptps{{\em Prog.\ Theoret.\ Phys.\ Suppl.\ }}
\def\PTP{{\em Prog.\ Theoret.\ Phys.\  }}

\def\JHEP{{\em JHEP} }
\def\JMP{{\em J. Math.\ Phys.} }
\def\NPB{{\em Nucl.\ Phys.} B}
\def\NP{{\em Nucl.\ Phys.} }
\def\PLB{{\em Phys.\ Lett.} B}
\def\PL{{\em Phys.\ Lett.} }
\def\PRL{\em Phys.\ Rev.\ Lett. }
\def\PRB{{\em Phys.\ Rev.} B}
\def\PRD{{\em Phys.\ Rev.} D}
\def\PR{{\em Phys.\ Rev.} }
\def\PRe{{\em Phys.\ Rep.} }
\def\AP{{\em Ann.\ Phys.\ (N.Y.)} }
\def\RMP{{\
em Rev.\ Mod.\ Phys.} }
\def\ZPC{{\em Z.\ Phys.} C}
\def\SCI{\em Science}
\def\CMP{\em Comm.\ Math.\ Phys. }
\def\MPLA{{\em Mod.\ Phys.\ Lett.} A}
\def\IJMPB{{\em Int.\ J.\ Mod.\ Phys.} B}
\def\cmp{{\em Com.\ Math.\ Phys.}}
\def\JPA{{\em J.\  Phys.} A}
\def\JPG{{\em J.\  Phys.} G}
\def\CQG{\em Class.\ Quant.\ Grav. }
\def\ATMP{{\em Adv.\ Theoret.\ Math.\ Phys.} }
\def\ibid{{\em ibid.} }
\vskip 1cm

\leftline{\bf References}

\renewenvironment{thebibliography}[1]
        {\begin{list}{[$\,$\arabic{enumi}$\,$]}  
        {\usecounter{enumi}\setlength{\parsep}{0pt}
         \setlength{\itemsep}{0pt}  \renewcommand{\baselinestretch}{1.2}
         \settowidth
        {\labelwidth}{#1 ~ ~}\sloppy}}{\end{list}}

\def\reftitle#1{}                


\begin{thebibliography}{99}
\small
\baselineskip=16pt

\bibitem{Witten1}
E.\ Witten, \jnl{\PRL}{38}{121}{1977}.

\bibitem{Fairlie1}
D.B.\ Fairlie, \jnl{\PLB}{82}{97}{1979};
\reftitle{Higgs' Fields And The Determination Of The Weinberg Angle}
\jnl{\JPG}{5}{L55}{1979}.
\reftitle{Two Consistent Calculations Of The Weinberg Angle}

\bibitem{Manton1}
P.\ Forgacs and N.\ Manton, \jnl{\CMP}{72}{15}{1980}.
\reftitle{Space-Time Symmetries In Gauge Theories}

N.\ Manton, \jnl{\NPB}{158}{141}{1979};
\reftitle{A New Six-Dimensional Approach To The Weinberg-Salam Model}


\bibitem{YH1}
Y.\ Hosotani, \jnl{\PLB}{126}{309}{1983}.

\bibitem{YH2}
Y.\ Hosotani, \jnl{\AP}{190}{233}{1989}.

\bibitem{Morris}
T.R.\ Morris, \jnl{\JHEP}{0501}{002}{2005}.

\bibitem{Takenaga1}
K.\ Takenaga,  \jnl{\PLB}{425}{114}{1998};
\reftitle{Supersymmetry Breaking through Boundary Conditions 
Associated with the   $U(1)_{R}$}
\jnl{\PRD}{58}{026004}{1998}.
\reftitle{Softly Broken Supersymmetric Gauge Theories through Compactifications}

\bibitem{Pomarol1}
A.\ Pomarol and M.\ Quiros, \jnl{\PLB}{438}{255}{1998}.
\reftitle{The Standard Model from extra dimensions}

\bibitem{Kawamura}
Y.~Kawamura, \jnl{\PTP}{103}{613}{2000}; 
\reftitle{Gauge Symmetry Reduction from the Extra Space $S^1/Z_2$}
\jnl{\PTP}{105}{999}{2001}.
\reftitle{Triplet-doublet Splitting, Proton Stability and Extra Dimension}

\bibitem{Antoniadis1}
I.\ Antoniadis, K.\ Benakli and M.\ Quiros,
\jnl{\it New. J.\ Phys.}{3}{20}{2001}.
\reftitle{Finite Higgs mass without Supersymmetry}

\bibitem{Hall1}
L.\ Hall and Y.\ Nomura, \jnl{\PRD}{64}{055003}{2001}; 
\reftitle{Gauge Unification in Higher Dimensions}

R.\ Barbieri, L.\ Hall and Y.\ Nomura,
\jnl{\PRD}{66}{045025}{2002};
\reftitle{Softly Broken Supersymmetric Desert from Orbifold Compactification}

A.\ Hebecker and J.\ March-Russell,
\jnl{\NPB}{625}{128}{2002};
\reftitle{The Structure of GUT Breaking by Orbifolding}


\bibitem{Lim1}
M.\ Kubo, C.S.\ Lim and H.\ Yamashita,
 \jnl{\MPLA}{17}{2249}{2002}.
\reftitle{The Hosotani Mechanism in Bulk Gauge Theories with an Orbifold Extra   Space $S^1/Z_2$}


\bibitem{HHHK}
N.\ Haba, M.\ Harada, Y.\ Hosotani and Y.\ Kawamura, 
\jnl{\NPB}{657}{169}{2003};   
{\it Erratum}, {\it ibid.}  B{\bf 669} (2003) {381}.
\reftitle{Dynamical Rearrangement of Gauge Symmetry on the Orbifold $S^1/Z_2$}

\bibitem{YH5}
Y.\ Hosotani,  in {\it ''Strong Coupling Gauge Theories and Effective Field
Theories"},  ed. M. Harada, Y. Kikukawa and K. Yamawaki (World Scientific
2003), p.\ 234. (hep-ph/0303066).
\reftitle{GUT on Orbifolds: Dynamical Rearrangement of Gauge Symmetry}

\bibitem{gaugeHiggs1}
G.\ Burdman and Y.\ Nomura, \jnl{\NPB}{656}{3}{2003}; 
\reftitle{Unification of Higgs and Gauge Fields in Five Dimensions}

C.\ Csaki, C.\ Grojean and H.\ Murayama, \jnl{\PRD}{67}{085012}{2003};
\reftitle{Standard Model Higgs From Higher Dimensional Gauge Fields}

I.\ Gogoladze, Y.\ Mimura and S.\ Nandi, 
\jnl{\PRL}{91}{141801}{2003};
\reftitle{Unity of Elementary Particles and Forces In Higher Dimensions}

C.A.\ Scrucca, M.\ Serone, L.\ Silvestrini and A.\ Wulzer,
\jnl{\JHEP}{0402}{49}{2004};
\reftitle{Gauge-Higgs Unification in Orbifold Models} 



\bibitem{HHK}
N.\ Haba,  Y.\ Hosotani and Y.\ Kawamura, 
\jnl{\PTP}{111}{265}{2004}.
\reftitle{Classification and dynamics of equivalence classes in SU(N) gauge theory   on the orbifold $S^1/Z_2$}

\bibitem{HNT1}
Y.\ Hosotani, S.\ Noda and K.\ Takenaga,
\jnl{\PRD}{69}{125014}{2004}.
\reftitle{Dynamical Gauge Symmetry Breaking and Mass Generation 
on the Orbifold $T^2/Z_2$}

\bibitem{HNT2}
Y.\ Hosotani, S.\ Noda and K.\ Takenaga,
\jnl{\PLB}{607}{276}{2005}.
\reftitle{Dynamical Gauge-Higgs Unification in the Electroweak Theory}

\bibitem{HHKY}
N.\ Haba,  Y.\ Hosotani,  Y.\ Kawamura and T.\ Yamashita, 
\jnl{\PRD}{70}{015010}{2004};
\reftitle{Dynamical symmetry breaking in Gauge-Higgs unification on orbifold}

N.\ Haba,  K.\ Takenaga, and T.\ Yamashita, 
hep-ph/0411250.
\reftitle{Higgs mass in the gauge-Higgs unification}

\bibitem{Quiros3}
M.\ Quiros,  in 
{\it ``Boulder 2002, Particle Physics and Cosmology''}, pages 549 - 601.
(hep-ph/0302189).
\reftitle{New Ideas in Symmetry Breaking}



\bibitem{RS}
L.\ Randall and R.\ Sundrum,  \jnl{\PRL}{83}{3370}{1999}.

\bibitem{Chang}
S.\ Chang, J.\ Hisano, H.\ Nakano, N.\ Okada and M.\ Yamaguchi, 
\jnl{\PRD}{62}{084025}{2000};
\reftitle{Bulk Standard Model in the Randall-Sundrum Background}

T.\ Gherghetta and A.\ Pomarol,
\jnl{\NPB}{586}{141}{2000}.
\reftitle{Bulk fields and supersymmetry in a slice of AdS}


\bibitem{Pomarol2}
R.\ Contino, Y.\ Nomura and A.\ Pomarol, \jnl{\NPB}{671}{148}{2003};
\reftitle{Higgs as a holographic pseudo-Goldstone boson}

K.\ Agashe, R.\ Contino and A.\ Pomarol,  hep-ph/0412089.
\reftitle{The minimal composite Higgs model}

\bibitem{Oda1}
K.\ Oda and A.\ Weiler, \jnl{\PLB}{606}{408}{2005}.
\reftitle{Wilson Lines in Warped Space: Dynamical Symmetry Breaking and Restoration}


\bibitem{Mabe}
Y.\ Hosotani and M.\ Mabe, hep-ph/0503020 (to appear in {\it Phys.\ Lett.} B).

\end{thebibliography}
\end{document}